\documentclass[12pt]{iopart}
\usepackage{iopams}
\usepackage{graphicx}
\graphicspath{{figures/}}
\usepackage{cite}
\usepackage{url}
\begin{document}

\title[A pure non-neutral plasma under an external harmonic field...]{A pure non-neutral plasma under an external harmonic field: equilibrium thermodynamics and chaos}

\author{B.F.I. Espinoza-Lozano, F.A. Calder\'on and L. Velazquez}

\address{Departmento de F\'isica, Universidad Cat\'olica del Norte, Av. Angamos 0610, Antofagasta, Chile.}
\ead{bel001@alumnos.ucn.cl, lvelazquez@ucn.cl, fcalderon@ucn.cl}
\begin{abstract}
Motivated by the precedent study of Ordenes-Huanca and Velazquez [JSTAT \textbf{093303} (2016)], we address the study of a simple model of a pure non-neutral plasma: a system of identical non-relativistic charged particles confined under an external harmonic field with frequency $\omega$. We perform the equilibrium thermo-statistical analysis in the framework of continuum approximation. This study reveals the existence of two asymptotic limits: the known Brillouin steady state at zero temperature, and the gas of harmonic oscillators in the limit of high temperatures. The non-extensive character of this model is evidenced by the associated thermodynamic limit, $N\rightarrow+\infty: U/N^{7/3}=const$, which coincides with the thermodynamic limit of a self-gravitating system of non-relativistic point particles in presence of Newtonian gravitation. Afterwards, the dynamics of this model is analyzed through numerical simulations. It is verified the agreement of thermo-statistical estimations and the temporal expectation values of the same macroscopic observables. The system chaoticity is addressed via numerical computation of Lyapunov exponents in the framework of the known \emph{tangent dynamics}. The temperature dependence of Lyapunov exponent $\lambda$ approaches to zero in the two asymptotic limits of this model, reaching its maximum during the transit between them. The chaos of the present model is very strong, since its rate is faster than the characteristic timescale of the microscopic dynamics $\tau_{dyn}=1/\omega$. A qualitative analysis suggests that such a strong chaoticity cannot be explained in terms of collision events because of their respective characteristic timescales are quite different,  $\tau_{ch}\propto \tau_{dyn}/N^{1/4}$ and $\tau_{coll}\propto \tau_{dyn}$.
\end{abstract}

\maketitle

\section{Introduction}

The study of dynamic and thermodynamic properties of systems with long-range interactions have received a great interest in the last decades. Paradigmatic examples of these systems are the non-neutral plasmas \cite{Malmberg1975,Malmberg1980,Davidson1990,ONeil1979,Driscoll1988,Driscoll1990,Gould1995,Huang1997,Dubin1999,Mattor2006,Fajans2003,Anderegg2009,Taylor1986,Huang1994,Boghosian1996,Anteneodo1997,Tsallis2001,Rodgers2009,Cabo2011a} and the astrophysical systems \cite{Lynden-Bell1999,Labini1998,Koyama2001,Torcini1999,Antonov1985,Binney2008,Lynden-Bell1967,Lynden-Bell1968,Michie1963,Michie1963b,King1962,King1965,King1966a,King1966b,Velazquez2003,Velazquez2009}. These systems exhibit macroscopic properties that differ from the ones observed in conventional short-range (extensive) systems of everyday applications of thermodynamics and statistical mechanics. The incidence of long-range interactions implies that these systems cannot be decomposed into independent subsystems, which means that they do not obey extensivity and additivity properties, neither the conventional extensive thermodynamic limit $N\rightarrow+\infty$ is applicable to them. Additionally, they can exhibit thermodynamic anomalies like \textit{negative heat capacities} and exotic phenomena like discontinuous microcanonical phase transitions, among others.

The dynamical evolution of these systems is also affected by the existence of long-range correlations among their constituents. This feature is analogous to the one exhibited by conventional extensive systems in conditions of criticality (e.g.: near a critical point). However, their existence is not associated with the occurrence of phase transitions, but the long-range character of their underlying interactions. Such long-range correlations can provoke dynamical anomalies such as slow relaxation and metastability. Therefore, most of practical realizations that involve systems with long-range interactions concern to \textit{out-of-equilibrium situations} \cite{Campa2009,Campa2014}. To make things worse, practical situations concerning to astrophysical systems and non-neutral plasmas are affected by external conditions that prevent their rigorous relaxation, such as the evaporation of constituents \cite{King1962,King1965,King1966a,King1966b,Velazquez2003,Velazquez2009}. In general, all possible internal or external conditions determining each practical situation of interest will provoke a considerable affectation in the whole macroscopic behavior of any system with long-range interaction, precisely, because of the existence of such long-range correlations. This feature differs from ordinary extensive systems, e.g.: the shape of a container does not modify the thermodynamic properties like specific heat or temperature of phase transitions.

The main interest of this contribution is to study the chaoticity of a \textit{pure non-neutral plasmas} (systems with only one class of identical charged particles like the electronic plasmas \cite{Huang1997,Dubin1999,Mattor2006,Fajans2003,Anderegg2009}) and its relation with the thermo-statistical description of these long-range interacting systems. This research was motivated by the recent work developed by Ordenes-Huanca and Velazquez \cite{Ordenes-Huanca2016}, which addressed the incidence of constituents evaporation on the thermodynamics of these systems. In this precedent work, the thermo-statistical description was developed by invoking a \textit{quasi-ergodicity} of microscopic dynamics along a quasi-stationary regime in presence of evaporation. The theoretical profiles predicted from this argument were compared with the experimental results of Huang and Driscoll \cite{Huang1994}. The good agreement observed in that study strongly suggested the relevance of these statistical arguments in a quasi-stationary regime reported in that experiment. Curiously, such an experimental data concerned to an electronic plasma where the incidence of collisional relation was negligible. Consequently, other relaxations mechanisms should be present in that experimental situation to justify the relevance of ergodicity.

According to \emph{chaotic hypothesis} \cite{GallavottiCohem:1995}, many-body nonlinear systems with strong chaotic properties must exhibit good statistical properties like the ones associated with ergodicity and mixing. Therefore, a plausible explanation for good fit of profiles of Huang-Driscoll experiment in the precedent study is the existence of a strong chaoticity in the microscopic dynamics of non-neutral plasmas. Pettini and co-workers reported in the past the existence of a very strong chaoticity in numerical simulations of a self-gravitating gas of particles driven by Newtonian forces \cite{Cipriani2003}. These authors demonstrated that the incidence of a collisional relaxation on the observed chaoticity was negligible, and they identify the mechanism of \emph{parametric resonance} as the main source of the observed instability \cite{Pettini1993}. The existing mathematical analogy among gravitation and Coulombic interactions suggests that a strong chaoticity should be also observed in the dynamics of pure non-neutral plasmas.

The paper is organized into sections as follows. The second section is devoted to present a simple model of pure non-neutral plasma, as well as the methodology for its study. We shall develop its thermo-statistical description in the framework of continuum approximation, as well as its dynamical description via numerical integration of Hamilton equations. Chaoticity itself will be studied via numerical computation of Lyapunov exponents in the framework of the called \textit{tangent dynamics}. Third and fourth sections will be devoted to present results and discussions. Final remarks and open problems are discussed in the fifth section.





\section{Methodology}

\subsection{The model: a pure non-neutral plasma trapped under an external harmonic field}
Experimentally, a non-neutral plasma can be effectively confined within external magnetic and electric fields of a Penning trap, whose Hamiltonian can be written as follows \cite{Dubin1999},
\begin{equation}
\label{eq:non-neutral-hamilton}
H(\mathbf{r},\mathbf{p})=\sum_i\frac{1}{2m}\left[\mathbf{p}_i-q\mathbf{A}(\mathbf{r}_i)\right]^2+ q\phi_T(\mathbf{r}_i)+\sum_{i<j}\frac{q^2}{|\mathbf{r}_i-\mathbf{r}_j|}.
\end{equation}
However, this type of confinement does not avoid the evaporation of constituents during the system dynamical evolution \cite{Ordenes-Huanca2016} neither the Hamiltonian (\ref{eq:non-neutral-hamilton}) obeys the standard form
\begin{equation}
\label{standardH}
H(q,p)=\sum_{i,j}\frac{1}{2}a^{ij}\left(q\right)p_{i}p_{j}+V(q),
\end{equation}
required for the application of Riemannian approach of Hamiltonian chaos \cite{Pettini1993}. Although this geometric framework will not be considered in this work, we think convenient to address a more simple Hamiltonian model:
\begin{equation}\label{eq:simplified-plasma-equation}
H(\mathbf{r},\mathbf{p})=\sum_{i}\frac{1}{2m}\mathbf{p}^{2}_{i}+\frac{1}{2}m\omega^{2}\mathbf{r}^{2}_{i}
+\sum_{i<j}\frac{q^{2}}{\left|\mathbf{r}_{i}-\mathbf{r}_{j}\right|},
\end{equation}
which avoids all previous difficulties, and it still enables us to study the chaoticity of a non-neutral plasma due to the long-range character of Coulombian forces. Hamiltonian (\ref{eq:simplified-plasma-equation}) represents a gas with $N$ identical non-relativistic point particles of mass $m$ and charge $q$, interacting through Coulomb's force under an external harmonic field with frequency $\omega$. This particular situation guarantees the fully confinement of the non-neutral plasma and avoids the incidence of evaporation. Consequently, the thermo-statistical description can be performed assuming a rigorous thermodynamic equilibrium.

\subsection{Thermo-statistical description}
The equilibrium one-body distribution function (DF) for the present model system is given by Maxwell-Boltzmann profile \cite{Reichl2016}:
\begin{equation}\label{MB}
    f(\mathbf{r},\mathbf{p})=A\exp\left[-\beta\varepsilon(\mathbf{r},\mathbf{p})\right].
\end{equation}
Here, $\varepsilon(\mathbf{r},\mathbf{p})$ denotes the mechanical energy of individual particles:
\begin{equation}\label{energia.perpart}
\varepsilon(\mathbf{r},\mathbf{p})=\frac{1}{2m}\mathbf{p}^{2}+q\varphi(\mathbf{r})+\frac{1}{2}m\omega^{2}\mathbf{r}^{2},
\end{equation}
$\beta=1/kT$ is the inverse temperature parameter, $k$ the Boltzmann constant, and $\varphi(\mathbf{r})$, the electrostatic potential. Particles distribution  $n(\mathbf{r})$ can be calculated as follows:
\begin{eqnarray}\label{densidad}
n(\mathbf{r})&=\int f(\mathbf{r},\mathbf{p})\frac{d^{3}\mathbf{p}}{(2\pi\hbar)^{3}} \\
&=\left(\frac{2m\pi}{\hbar^{2}\beta}\right)^{\frac{3}{2}}A
     \exp\left\{-\beta\left[\frac{1}{2}m\omega^{2}\mathbf{r}^{2}+q\varphi(\mathbf{r})\right]\right\},\nonumber
\end{eqnarray}
which obeys the normalization condition:
\begin{equation}\label{norma}
N[f]= \int f(\mathbf{r},\mathbf{p})\frac{d^{3}\mathbf{r} d^{3}\mathbf{p}}{(2\pi\hbar)^{3}}=\int n(\mathbf{r})d^{3}\mathbf{r}=N.
\end{equation}
The total energy $U$ is given by:
\begin{equation}\label{tot.energia}
 U=\frac{3N}{2\beta}+\int\frac{1}{2}m\omega^{2}\mathbf{r}^{2}n(\mathbf{r})d^{3}\mathbf{r}+\frac{1}{2}\int q\varphi(\mathbf{r}) n(\mathbf{r})d^{3}\mathbf{r},
\end{equation}
where the first term represents the kinetic energy:
\begin{equation}
    \int \frac{\mathbf{p}^{2}}{2m}f(\mathbf{r},\mathbf{p})\frac{d^{3}\mathbf{r}d^{3}\mathbf{p}}{(2\pi\hbar)^{3}}=\int \frac{3}{2\beta}n(\mathbf{r})d^{3}\mathbf{r}=\frac{3}{2}\frac{N}{\beta}.
\end{equation}

The electrostatic potential $\varphi(\mathbf{r})$ is related to the particles density $n(\mathbf{r})$ throughout Poisson equation, $\Delta\varphi(\mathbf{r})=-4\pi qn(\mathbf{r})$. Introducing the effective field $u(\mathbf{r})$:
\begin{equation}
u(\mathbf{r})=\frac{1}{2}m\omega^{2}\mathbf{r}^{2}+q\varphi(\mathbf{r}),
\end{equation}
and the dimensionless potential  $\Phi(r)$:
\begin{equation}
\Phi(\mathbf{\mathbf{r}})=\beta\left[u(0)-u(\mathbf{r})\right],
\end{equation}
the particles density can be rewritten as:
\begin{equation}\label{den.D}
n(\mathbf{r})=n_{0}\exp\left[\Phi(\mathbf{r})\right],
\end{equation}
where $n_{0}$ is the central density:
\begin{equation}\label{CC}
n_{0}=\left(\frac{2m\pi}{\hbar^{2}\beta}\right)^{\frac{3}{2}}A\exp\left[-\beta u(0)\right].
\end{equation}
Additionally, it is convenient to introduce the characteristic radius constant $r_{c}$:
\begin{equation}\label{rc}
\frac{1}{r_{c}^{2}}=q^{2}n_{0}\beta,
\end{equation}
the dimensionless radius variable $\xi=r/r_{c}$, and the auxiliary constant $\lambda$:
\begin{equation}\label{lambda}
\lambda=\frac{3m\beta\omega^{2}}{4\pi}r_{c}^{2}=\frac{3m\omega^{2}}{4\pi q^{2}n_{0}}.
\end{equation}
The above definitions enable us to rephrase Poisson equation with spherical symmetry as follows:
\begin{equation}\label{pb.potadim}
\frac{1}{\xi^{2}}\frac{d}{d\xi}\left[\xi^{2}\frac{d}{d\xi}\Phi(\xi)\right]=4\pi\left[e^{\Phi(\xi)}-\lambda\right],
\end{equation}
which should be solved by numerical integration using the following boundary conditions at the origin:
\begin{equation}\label{frontera.origen}
\Phi(0)=0,\: \frac{d}{d\xi}\Phi(0)=0.
\end{equation}
Additionally, it is necessary to take into account the asymptotic behavior of the dimensionless potential for large distances. For spherical solutions, the electrostatic potential can be expressed as follows:
\begin{equation}\label{sol.pot}
\frac{d}{dr}\varphi(r)=-\frac{qN(r)}{r^{2}},
\end{equation}
where $N(r)$ is the number of particles enclosed inside a sphere of radius $r$. Considering the first derivative of the dimensionless potential:
\begin{equation}\label{dphi}
  \frac{d\Phi(\xi)}{d\xi}=\beta r_{c}\left[m\omega^{2}r+q\frac{d\varphi(r)}{dr}\right],
\end{equation}
and introducing the dimensionless parameter $\eta$:
\begin{equation}\label{eta}
\eta=\frac{q^{2}N\beta}{r_{c}},
\end{equation}
one obtains the following asymptotic expression for $\eta$:
\begin{equation}
\eta=\lim_{\xi\rightarrow+\infty}\left[\frac{4\pi}{3}\lambda\xi^{3}+\xi^{2}\frac{d}{d\xi}\Phi(\xi)\right].
\end{equation}

At zero temperature, the pure non-neutral plasma in a magnetic trap adopts the called \textit{Brillouin steady state} \cite{Gould1995}, a profile of uniform density where electrostatic repulsion forces are exactly compensated with electric and magnetic fields of the Penning trap. An analogous situation also appears in the zero temperature limit for the present model. Considering the vanishing of the resulting force when $T\rightarrow 0$:
\begin{equation}
  \mathbf{F}(\mathbf{r})=q\mathbf{E}(\mathbf{r})-m\omega^{2}\,\mathbf{r}=0,
\end{equation}
one can apply the divergence to obtain the Brillouin density:
\begin{equation}\label{nb}
 n_{B}=\frac{3m\omega^{2}}{4\pi q^{2}}
\end{equation}
associated with the presence of the external harmonic field with frequency $\omega$. This profile of constant density is not infinitely extended because of the number of particles $N$ is bound. The right expression is given by:
\begin{equation}\label{perfil.Brillouin}
n(r)=\left\{
\begin{array}{cc}
n_{B}, & \mbox{ if }r<R_{B}, \\
0, & \mbox{ if }r\geq R_{B}, \\
\end{array}
\right.
\end{equation}
where $R_{B}$ is the Brillouin radius:
\begin{equation}
n_{B}\frac{4\pi}{3} R^{3}_{B}=N,
\end{equation}
which defines the linear size of the system at zero temperature.

For the sake of convenience, we shall refer numerical results of this study by using a set of characteristic units associated with Brillouin state. The linear distances and particles densities will be referred into units of Brillouin radius $R_{B}$ and density $n_{B}$. Using these units, the particles density  (\ref{den.D}) can be rewritten as:
\begin{equation}\label{den.adim.B}
\bar{n}(\bar{r})=\frac{1}{\lambda}\exp\left[\Phi(\bar{r})\right],
\end{equation}
where $\bar{n}(r)=n(r)/n_{B}$ and $\bar{r}=r/R_{B}=\xi r_{c}/R_{B}$. It was considered here the relation:
\begin{equation}\label{lambda.n0}
\frac{n_{0}}{n_{B}}=\frac{1}{\lambda},
\end{equation}
which is derived from definition (\ref{lambda}). According to Eq.(\ref{den.adim.B}), the Brillouin steady state appears in the limit $\lambda\rightarrow 1$, where $\bar{n}(\bar{r})=1$ and $\Phi(\bar{r})=0$. Additionally, one can introduce the characteristic units:
\begin{equation}
 U_{B}=\frac{q^{2}N^{2}}{R_{B}}\mbox{ and } T_{B}=\frac{q^{2}N}{R_{B}k}
\end{equation}
for the energy and the temperature, respectively. The inverse temperature $\beta$ and the characteristic radius $r_{c}$ defined in Eq.(\ref{rc}) can be rewritten into Brillouin units as:
 \begin{equation}
\bar{\beta}=\frac{T_{B}}{T}=\left(\frac{4\pi}{3}\lambda\eta^{2}\right)^{\frac{1}{3}}\mbox{ and }\bar{r}_{c}=\frac{r_{c}}{R_{B}}=\left(\frac{4\pi}{3}\frac{\lambda}{\eta}\right)^{\frac{1}{3}}.
 \end{equation}
Let us now obtain the working expressions for the total energy (\ref{tot.energia}) in Brillouin units. The kinetic energy contribution is expressed as:
\begin{equation}\label{contr.cinet}
\bar{K}=\frac{K}{U_{B}}=\frac{3}{2}\frac{1}{\bar{\beta}}.
\end{equation}
The total potential energy associated with the external harmonic field:
\begin{equation}\label{contri.arm.ext}
\frac{W}{U_{B}}=\frac{1}{2}\frac{\bar{r}_{c}^{3}}{\bar{\beta}}\mathbb{Q},
\end{equation}
where $\mathbb{Q}$ denotes the auxiliary integral:
\begin{equation}
\mathbb{Q}=\int^{+\infty}_{0}\xi^{2}\exp[{\Phi(\xi)}]4\pi\xi^{2}d\xi.
\end{equation}
Finally, let us obtain the expression for the total electrostatic energy. Rephrasing this contribution in terms of dimensionless potential $\Phi(\mathbf{r})$:
\begin{equation}\label{contr.part}
V=\frac{1}{2}Nq\varphi(0)-\frac{1}{2}W-\int\frac{1}{2\beta}\Phi(\mathbf{r})n(\mathbf{r})d^{3}\mathbf{r}
\end{equation}
and rewriting it into Brillouin energy units $U_{B}$, one obtains:
\begin{equation}\label{contr.part2}
\bar{V}=\frac{V}{U_{B}}= \frac{3}{8\pi}\frac{\bar{r}_{c}^{2}}{\lambda}\mathbb{C}-\frac{1}{4}\frac{\bar{r}_{c}^{3}}{\bar{\beta}}\mathbb{Q}-
    \frac{3}{8\pi}\frac{\bar{r}_{c}^{3}}{\lambda\bar{\beta}}\mathbb{P},
\end{equation}
where $\mathbb{C}$ and $\mathbb{P}$ are the following auxiliary integrals:
\begin{eqnarray}\label{imtegr.func}
 \mathbb{C}&=&\int^{+\infty}_{0}\exp[{\Phi(\xi)}]4\pi\xi d\xi,\\
 \mathbb{P}&=&\int^{+\infty}_{0}\Phi(\xi)\exp[{\Phi(\xi)}]4\pi\xi^{2}d\xi.
 \end{eqnarray}
 The sum of all these contributions is:
 \begin{equation}
\bar{U}=\frac{U}{U_{B}}=\frac{3}{2}\frac{1}{\bar{\beta}}+\frac{1}{4}\frac{\bar{r}_{c}^{3}}{\bar{\beta}}\mathbb{Q}+
    \frac{3}{8\pi}\frac{\bar{r}_{c}^{2}}{\lambda}\left(\mathbb{C}-\frac{\bar{r}_{c}}{\bar{\beta}}\mathbb{P}\right).
\end{equation}

\subsection{Dynamical description}

Hamiltonian equations for the model (\ref{eq:simplified-plasma-equation}) are given by:
\begin{equation}\label{H.plasma}
\dot{\mathbf{r}}_{i}=\frac{1}{m}\mathbf{p}_{i},\; \dot{\mathbf{p}}_{i}=
-m\omega^{2}\mathbf{r}_{i}+\sum_{j\neq i} \frac{q^{2}}{\left|\mathbf{r}_{ij}\right|^{3}}\mathbf{r}_{ij},
\end{equation}
where $\mathbf{r}_{ij}=\mathbf{r}_{i}-\mathbf{r}_{j}$ is the separation vector oriented from $j$-th to $i$-th particle. Numerical integration of this conservative system is better performed using some sympletic algorithm \cite{Forest1990,Yoshida1990}. An efficient and precise sympletic algorithm was proposed by Casetti \cite{Casetti1993}:
\begin{eqnarray}\label{simpletico}
    \tilde{\mathbf{r}}_{i}=\mathbf{r}_{i}(t),\: \tilde{\mathbf{p}}_{i}=\mathbf{p}_{i}(t)-\frac{1}{2}\Delta t\frac{\partial}{\partial \tilde{\mathbf{r}}_{i}}V(\tilde{\mathbf{r}}),\\ \nonumber
    \mathbf{r}_{i}(t+\Delta t)= \tilde{\mathbf{r}}_{i}+\Delta t\frac{1}{m}\tilde{\mathbf{p}}_{i},\: \mathbf{p}_{i}(t+\Delta t)= \tilde{\mathbf{p}}_{i}-\frac{1}{2}\Delta t\frac{\partial}{\partial{\mathbf{r}}_{i}}V\left[\mathbf{r}(t+\Delta t)\right],\\ \nonumber
    \hat{\mathbf{p}}_{i}=\mathbf{p}_{i}(t+\Delta t),\:\hat{\mathbf{r}}_{i}=\mathbf{r}_{i}(t+\Delta t)+\frac{1}{2m}\Delta t\hat{\mathbf{p}}_{i},\\ \nonumber
    \mathbf{p}_{i}(t+2\Delta t)= \hat{\mathbf{p}}_{i}-\Delta t\frac{\partial}{\partial \hat{\mathbf{r}}_{i}}V\left[\hat{\mathbf{r}}\right],\:
    \mathbf{r}_{i}(t+2\Delta t)=\hat{\mathbf{r}}_{i}+\frac{1}{2m}\Delta t \mathbf{p}_{i}(t+2\Delta t),
\end{eqnarray}
which was employed by Pettini and co-workers in the astrophysical situation \cite{Pettini1993,Cerruti-Sola1995a}. This same algorithm will be considered in this work.

The instability of trajectories can be studied using the tangent dynamics, that is, the linearization of separation dynamics of two trajectories that are infinitely close:
\begin{equation}\label{tangent_dyn}
 \dot{\boldsymbol{ \xi}}_{i}=\frac{1}{m}\boldsymbol{\eta}_{i},\:    \dot{\boldsymbol{\eta}}_{i}=-m\omega^{2}\boldsymbol{\xi}_{i}+\sum_{j\neq i} \frac{q^{2}}{\left|\mathbf{r}_{ij}\right|^{3}}\left\{\boldsymbol{\xi}_{ij}-
 3\left(\mathbf{n}_{ij}\cdot\boldsymbol{\xi}_{ij}\right)\mathbf{n}_{ij}\right\},
\end{equation}
where $\mathbf{n}_{ij}$ and $\boldsymbol{\xi}_{ij}$ are given by:
\begin{equation}
\mathbf{n}_{ij}=\frac{\mathbf{r}_{ij}}{|\mathbf{r}_{ij}|}\mbox{ and }\boldsymbol{\xi}_{ij}=\boldsymbol{\xi}_{i}-\boldsymbol{\xi}_{j}.
\end{equation}
Numerical integration of the above equations is coupled to Hamilton equations (\ref{H.plasma}). For the numerical integration of the tangent dynamics, we shall employ a second-order Euler scheme:
\begin{eqnarray}\nonumber
\boldsymbol{\xi}_{i}\left(t+\Delta t\right)&=&\boldsymbol{\xi}_{i}\left(t\right)+\frac{1}{m}\boldsymbol{\eta}_{i}\left(t\right)\Delta t+\frac{1}{2m}\dot{\boldsymbol{ \eta}}_{i}\left(t\right)\Delta t^{2}+\frac{1}{6m}\ddot{\boldsymbol{ \eta}}_{i}\left(t\right)\Delta t^{3},\\
\boldsymbol{\eta}_{i}\left(t+\Delta t\right)&=&\boldsymbol{\eta}_{i}\left(t\right)+\dot{\boldsymbol{\eta}}_{i}\left(t\right)\Delta t+\frac{1}{2m}\ddot{\boldsymbol{ \eta}}_{i}\left(t\right)\Delta t^{2},
\end{eqnarray}
which requires the second time derivative $\ddot{\boldsymbol{\eta}}_{i}$ of speed separation $\boldsymbol{\eta}_{i}$:
\begin{eqnarray}\nonumber
\ddot{\boldsymbol{\eta}}_{i}&=-{\omega}^2{\boldsymbol{\eta}}_{i}+\sum_{j\neq i} \frac{q^{2}}{\left|\mathbf{r}_{ij}\right|^{3}}\left[-3(\mathbf{n}_{ij}\cdot\boldsymbol{\kappa}_{ij})\left\{\boldsymbol{\xi}_{ij}
-3\left(\mathbf{n}_{ij}\cdot\boldsymbol{\xi}_{ij}\right)\mathbf{n}_{ij}\right\}+\dot{\boldsymbol{\xi}}_{ij}\right. \\                          &\qquad\qquad\left.-3\left(\mathbf{n}_{ij}\cdot\dot{\boldsymbol{\xi}}_{ij}\right)\mathbf{n}_{ij}-3\left(\dot{\mathbf{n}}_{ij}
\cdot\boldsymbol{\xi}_{ij}\right)\mathbf{n}_{ij}-3\left(\mathbf{n}_{ij}
\cdot\boldsymbol{\xi}_{ij}\right)\dot{\mathbf{n}}_{ij}\right].
\end{eqnarray}
For the sake of briefly, we have considered here the following notations:
\begin{equation}
\boldsymbol{\kappa}_{ij}=\frac{\mathbf{v}_{ij}}{|\mathbf{r}_{ij}|},\:\mathbf{v}_{ij}=\mathbf{v}_{i}-\mathbf{v}_{j}
\mbox{ and }\dot{\mathbf{n}}_{ij}=\left\{\boldsymbol{\kappa}_{ij}-(\mathbf{n}_{ij}\cdot \boldsymbol{\kappa}_{ij})\mathbf{n}_{ij}\right\}.
\end{equation}

Finally, the calculation of Lyapunov exponent is performed considering the limit:
\begin{equation}\label{Lyapunov.limit}
\lambda=\lim_{t\rightarrow+\infty}\frac{1}{t}\ln\frac{\left\|\Delta x(t)\right\|}{\left\|\Delta x(0)\right\|},
\end{equation}
where the norm of deviation of trajectories $\left\|\Delta x(t)\right\|$ is defined as follows:
\begin{equation}\label{norma.desviacion}
\left\|\Delta x(t)\right\|^{2}=\sum_{i}\frac{1}{2m}\boldsymbol{\eta}^{2}_{i}(t) +\frac{1}{2}m\omega^{2}\boldsymbol{\xi}^{2}_{i}(t).
\end{equation}
For a computational viewpoint, the calculation the limit (\ref{Lyapunov.limit}) can be rephrased as the average of a discrete time series with period $\tau$. Denoting the  $n$-th time instant $t_{n}=n \tau$ and the corresponding exponential dispersion $\hat{\lambda}_{n}$ as:
\begin{equation}
\hat{\lambda}_{n}=\frac{1}{\left(t_{n}-t_{n-1}\right)}\ln\frac{\left\|\Delta x(t_{n})\right\|}{\left\|\Delta x(t_{n-1})\right\|},
\end{equation}
the momentum of $m$-th order $\lambda^{(m)}_{n}$ of the exponential dispersion $\hat{\lambda}_{n}$ can be expressed as follows:
\begin{equation}
\lambda^{(m)}_{n}=\lambda^{(m)}_{n-1}+\frac{1}{n}\left[\left(\hat{\lambda}_{n}\right)^{m}-\lambda^{(m)}_{n-1}\right].
\end{equation}
Accordingly, the infinite time limit (\ref{Lyapunov.limit}) is now replaced by its discrete version as:
\begin{equation}\label{discrete.Lyap}
\lambda=\lim_{n\rightarrow+\infty}\lambda^{(1)}_{n}.
\end{equation}
Of course, the Lyapunov exponent $\lambda$ can be estimated by the $n$-th value $\lambda^{(1)}_{n}$. Its statistical error $\delta \lambda_{n}$ can be expressed in terms of the square dispersion $\sigma^{2}_{n}=\lambda^{(2)}_{n}-\left(\lambda^{(1)}_{n}\right)^{2}$ as follows:
\begin{equation}
\delta \lambda_{n}=\sigma_{n}/\sqrt{n\tau/\tau_{d}},
\end{equation}
with $\tau_{d}$ being the \emph{decorrelation time} (the necessary time interval that the dynamics produces a statistical independent microscopic configuration). Taking into consideration the physical meaning of Lyapunov exponent $\lambda$, one could estimate the decorrelation time $\tau_{d}$ using the \emph{chaotization time}, $\tau_{d}\sim\tau_{ch}\sim 1/\lambda$. This consideration enable us to estimate the relative error as:
\begin{equation}
\frac{\delta \lambda_{n}}{\lambda^{(1)}_{n}}=\frac{\sigma_{n}}{\lambda^{(1)}_{n}}\frac{1}{\sqrt{n\lambda^{(1)}_{n}\tau}}.
\end{equation}
For a general physical observable $O$, e.g.: the total kinetic energy $K$, the calculation of its temporal expectation can be implemented using discrete time series as follows:
\begin{equation}
O_{n}=O_{n-1}+\frac{1}{n}\left(\hat{O}_{n}-O_{n-1}\right),
\end{equation}
where $\hat{O}_{n}=O\left(t_{n}\right)$ is its corresponding value at the time instant $t_{n}$.

\section{Results of the thermodynamic description}

\subsection{Thermodynamic limit and non-extensive character}

The short-range interacting systems obey the extensive thermodynamic limit $N\rightarrow\infty$:
\begin{equation}
 N\rightarrow\infty\Rightarrow \frac{S}{N}=\emph{const}, \,\frac{U}{N}=\emph{const},\, \frac{V}{N}=\emph{const},
\end{equation}
which guarantees the intensive character of some relevant observables and thermodynamic parameters like temperature and particles density. For systems with long-range interactions, the additivity and separability of extensive systems are non necessarily applicable. This type of arguments should be considered with care for each particular situation. For the concrete case of astrophysical system composed of non-relativistic point particles that interact among them through Newtonian gravitation, it has been claimed the relevance of the following thermodynamic limit \cite{Velazquez2016}:
\begin{equation}
  N\rightarrow\infty\Rightarrow \frac{S}{N}=\emph{const}, \,\frac{U}{N^{\frac{7}{3}}}=\emph{const},\, VN=\emph{const},
\end{equation}
which accounts for a non-extensive behavior. One should expect a similar thermodynamic limit for the case of a pure non-neutral plasma considered in this study. In order to check this possibility, let us consider the arguments already employed in the astrophysical case. Considering the characteristic linear dimension of the present situation as the Brillouin radius $r_{c}\sim R_{B}$, as well as the characteristic momentum $p_{c}$ derived from the Brillouin energy $U_{B}=q^{2}N^{2}/R_{B}$:
\begin{equation}
  p_{c}=\sqrt{\frac{2mU_{B}}{N}},
\end{equation}
these quantities can be employed to estimate the volume of phase space $\Omega$ as:
\begin{equation}
  \Omega\sim\frac{1}{N!}\left(\frac{p_{c}r_{c}}{\hbar}\right)^{3N},
\end{equation}
which leads to the following estimation for the entropy $S=k\ln\Omega$:
\begin{equation}\label{entrop}
  S\sim\frac{Nk}{2}\ln\left(\frac{8m^{3}q^{6}NR_{B}^{3}e^{2}}{\hbar^{6}}\right).
\end{equation}
According to the expression (\ref{entrop}), the entropy per particle $S/N$ remains finite when $N\rightarrow\infty$ after assuming the non-extensive scaling laws:
\begin{equation}\label{next.ThL}
    U/N^{7/3}=const \mbox{  and } RN^{1/3}=const.
\end{equation}
This limit evidences the non-extensive character of the pure non-neutral plasma described by the Hamiltonian (\ref{eq:non-neutral-hamilton}), which is essentially the same behavior associated with astrophysical case. Considering the scaling behavior of the particles density $n$, one can obtain the scaling behavior of the frequency $\omega$ of the external harmonic field
\begin{equation}\label{omega.N.ThL}
n/N^{2}=const\Rightarrow\omega/N=const,
\end{equation}
where the expression of Brillouin density (\ref{nb}) was taken into consideration.

\subsection{Equilibrium thermodynamics}

\begin{figure}[th!]
\centering
\includegraphics[width=4.5in]{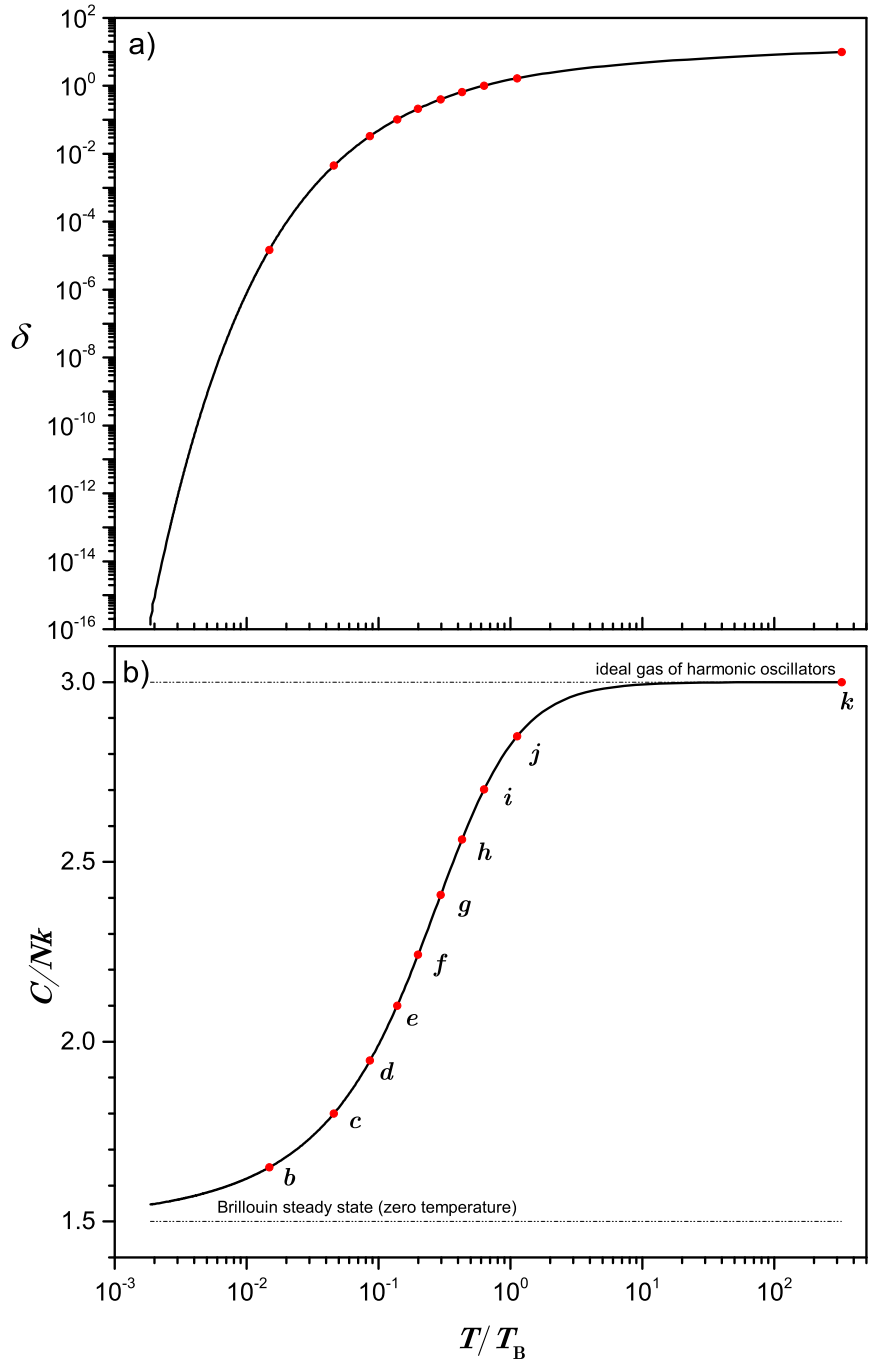}
\caption{Temperature dependencies of some relevant observables. Upper panel: The function $\delta=\ln\left(n_{B}/n_{0}\right)$, where $n_{0}$ is the central density, and $n_{B}=3m\omega^{2}/4\pi q^{2}$, the Brillouin density. Bottom panel: Specific heat capacity per particle $C/N=\left(dU/dT\right)/N$ (in units of Boltzmann constant $k$). With the growth of temperature, the system undergoes a transition from a low temperature limit (the Brillouin steady state) towards a high temperature limit (the ideal gas of harmonic oscillators).}\label{g01_limits.eps}
\end{figure}

\begin{figure}[t!]
\centering
\includegraphics[width=4.5in]{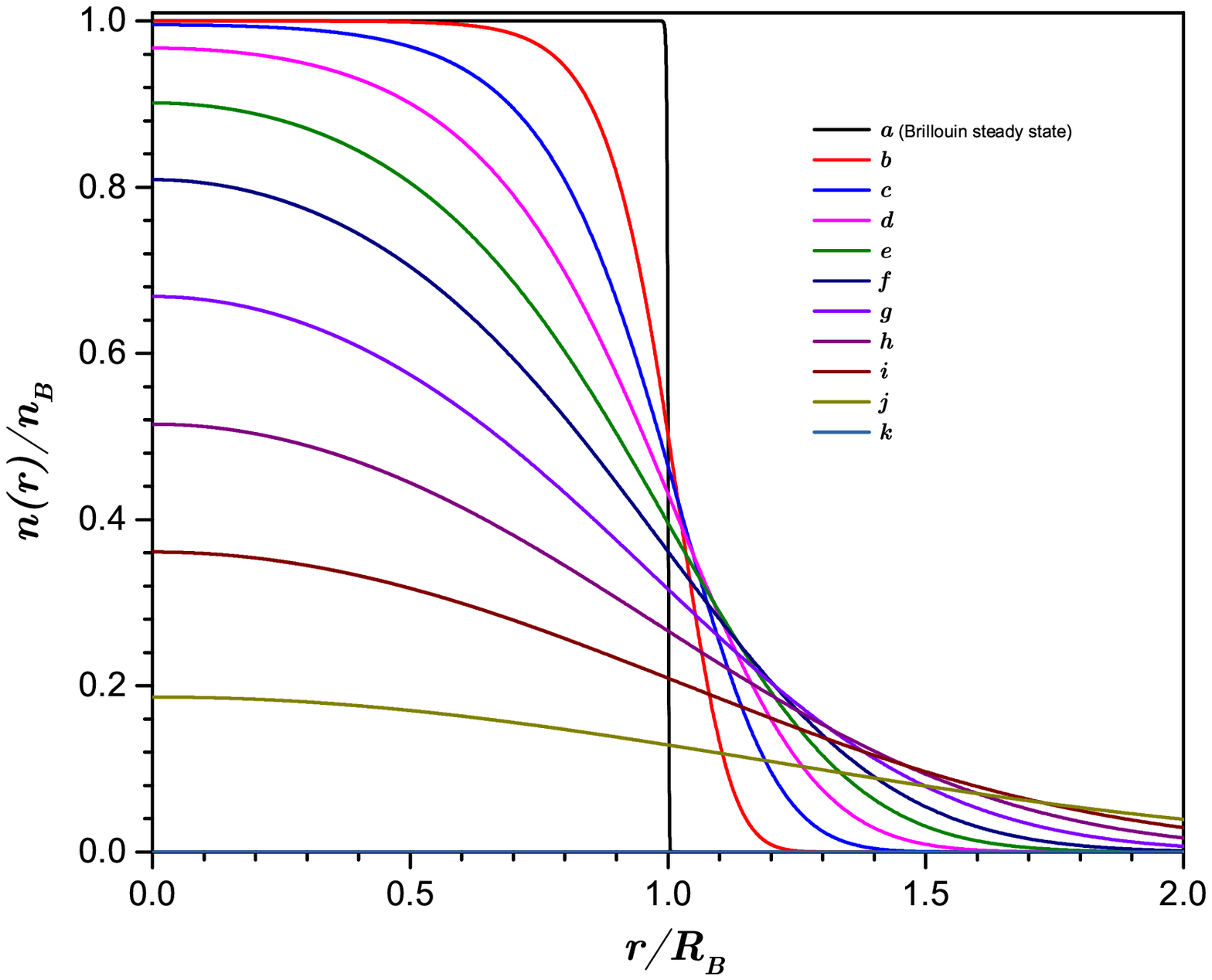}
\caption{Density radial distributions for some characteristic values of the temperature [the red point highlighted in figure \ref{g01_limits.eps}]. One observes here how the radial distributions approach the Brillouin profile (\ref{perfil.Brillouin}) as the temperature approaches to zero.}\label{g02_profiles.eps}
\end{figure}

Since the number of particles of the model system under consideration is finite, the particles density (\ref{den.D}) must be a bound function. For large distances, the particles density should be a monotonous decreasing function. Analysing the Poisson problem (\ref{pb.potadim}), the behaviour of the density profile, or even the one associated with the dimensionless potential $\Phi(\xi)$, is driven by the sign of the function  $(\exp\left[\Phi(\xi)\right]-\lambda)$. If this function is positive definite, the dimensionless potential will be a monotonous increasing function, and hence, this possibility cannot describe a density profile with a finite number of particles. Therefore, this function has to be negative definite in order to describe a monotonous decreasing function for the dimensionless $\Phi(\xi)$. The maximum will always take place at the origin $\xi=0$, so that, the integration parameter $\lambda\geq 1$. The lower bound $\lambda=1$ corresponds to the Brillouin steady state, that is, the zero temperature limit. For the sake of convenience, let us introduce the auxiliary parameter $\delta=\ln(\lambda)=\ln\left(n_{B}/n_{0}\right)$, which is a measure of the central density $n_{0}$ in units of the Brillouin density $n_{B}$.

We show in figure \ref{g01_limits.eps} the temperature dependencies (in units of the Brillouin temperature $T_{B}$) of two relevant observables, the auxiliary parameter $\delta$ and the specific heat (in units of the Boltzmann constant $k$). According to these results, the auxiliary parameter $\lambda$ exhibits a wide range of values to describe the physics associated with the present situation. By itself, the mathematical behavior of the parameter $\delta$ evidences difficulties of numerical integration of Poisson problem (\ref{pb.potadim}) in the low temperature region. On the other hand, temperature dependence of the specific heat (per particles) evidences the existence of two asymptotic limits: the zero temperature limit where appears the Brillouin steady state, and the high temperature limit, where the system behaves as a ideal gas of harmonic oscillators.

In the low temperature limit, we observe that the specific heat approaches to the asymptotic value $C/Nk=3/2$, which corresponds to the heat capacity of an ideal gas at constant volume $V_{B}=4\pi R^{3}_{B}/3$. Such a capacity is explained by the contribution of kinetic degrees of freedom only. The repulsive electrostatic forces among the charged particles is fully compensated with the linear forces of the external harmonic field. The contribution of spatial degrees of freedom to the heat capacity is vanishing at zero temperature. In the limit of high temperatures, the density of the system decreases so much that electrostatic potential energy is almost negligible, so that, the system behaves here as an ideal gas of harmonic oscillators. According to the equipartition theorem, the specific heat capacity asymptotically approaches the value  $C/Nk=3$. During the transition from the low to high temperature limits, one observes an effective unfrozen of the oscillatory degrees of freedom. It is worthy to comment that the present behavior cannot be associated to the occurrence of a phase transition. Actually, it is rather analogous to the unfrozen of oscillatory and vibrational degrees of freedom in gases and solids due to quantum effects \cite{Reichl2016}.

\begin{figure}[t!]
\centering
\includegraphics[width=6.0in]{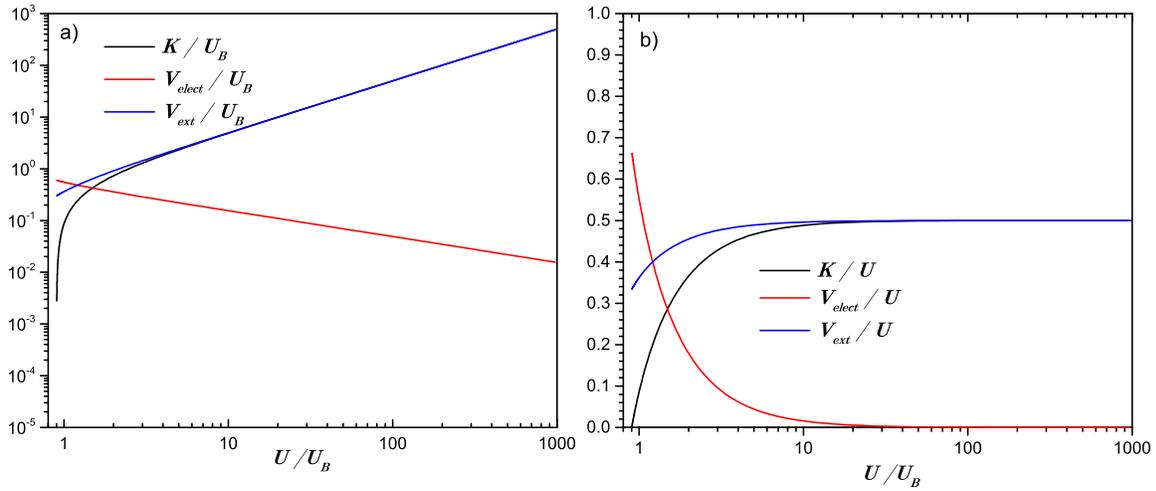}
\caption{Microcanonical dependencies of three contributions of the total energy $U=K+V_{ext}+V_{elect}$: the total kinetic energy $K$, the potential energy of the harmonic field $V_{ext}$ and the total electrostatic energy $V_{elect}$. Left panel: Their absolute values. Right panel: Their relative values (their ratios with the total energy $U$). Considering the high temperature limit, all these dependencies account for the transition towards the thermodynamic behavior associated with the ideal gas of harmonic oscillators, e.g.: the equalization between the total kinetic energy $K$ and the total potential energy $V_{elect}$ of the harmonic field, and the vanishing of the total electrostatic energy $V_{elect}$. The indicators that account for the system approaching towards Brillouin steady state are more subtle: the potential total energies of the harmonic field $V_{ext}$ and the electrostatic forces $V_{elect}$ are non-vanishing at zero temperature.}\label{g03_dyn_obs.eps}
\end{figure}

Let us discuss now some other relevant thermodynamic observables. We have highlighted in figure \ref{g01_limits.eps} a series of states (the red points labelled with Latin letters $a-k$). The same ones were uniformly located between the two asymptotic values of the specific heat capacity. For these points, we have obtained the radial distribution profiles $ n\left(r\right)$ shown in figure \ref{g02_profiles.eps}. The step function (\ref{perfil.Brillouin}) associated to the Brillouin state is established when the temperature $T$ approaches to zero. For nonzero temperatures, the radial distribution profiles turn smooth functions (without discontinuities), which decreases monotonically with the growth of radial coordinate $r$. The associated central density $n_{0}$ of these profiles decreases with the growth of temperature (this behavior is also evidenced by the temperature dependence of the quantity $\delta=\ln(n_{B}/n_{0})$ shown in figure \ref{g01_limits.eps}). At the limit of hight temperatures $T\gg T_{B}$, the radial profiles turn the Gaussian profile:
\begin{equation}\label{Gaussian.profile}
n_{G}\left(r\right)=\left(\frac{m\omega^{2}}{4\pi kT}\right)^{3/2}N\exp\left[-m\omega^{2}r^{2}/2kT\right]
\end{equation}
associated to an ideal gas of harmonic oscillators. Accordingly, the central density $n_{0}$ adopts in this asymptotic limit the analytical form $n_{0}=n_{B}\left(T_{B}/T\right)^{3/2}/6\sqrt{\pi}$. We show in figure \ref{g03_dyn_obs.eps} the dependencies of three contributions of the total energy $U=K+V_{ext}+V_{elect}$ (the total kinetic energy $K$, the potential energy of the harmonic field $V_{ext}$ and the total electrostatic energy $V_{elect}$) \emph{versus} the total energy $U$. For low energies, the kinetic energy drops to zero, while both potential energy contributions exhibit finite values due to the system adopts the Brillouin profile (\ref{perfil.Brillouin}).  For low energies, the electrostatic potential energy approaches to zero and there is an asymptotic equipartition between the kinetic energy and the potential energy of the external harmonic field.

\subsection{Comparison with astrophysical systems}

\begin{figure}[t!]
\centering
\includegraphics[width=4.5in]{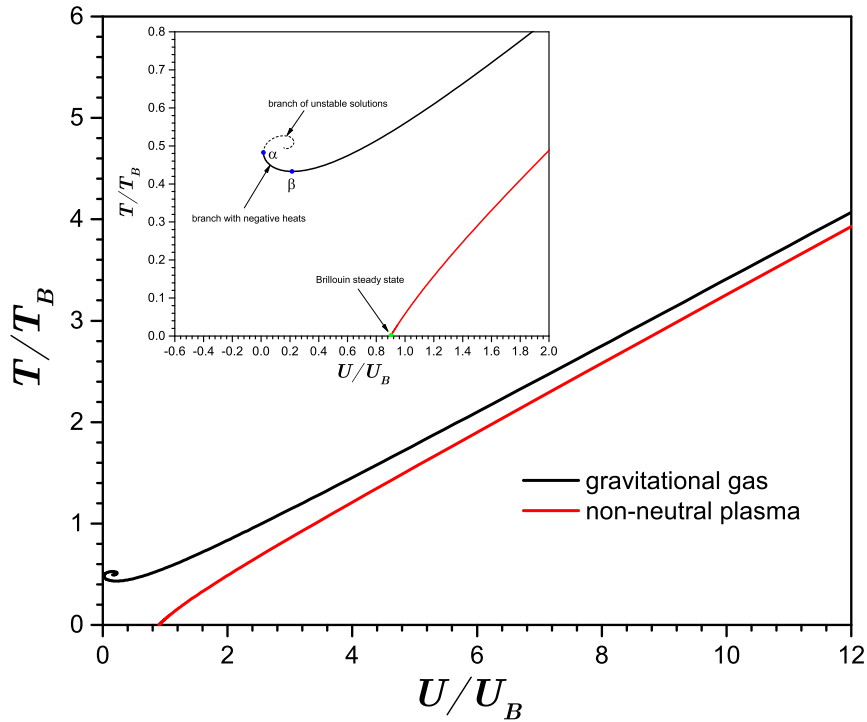}
\caption{Comparison between the microcanonical caloric curves corresponding to the non-neutral plasma model (\ref{eq:non-neutral-hamilton}) and the gravitational gas (\ref{gravitational.model}). Both cases exhibit the same asymptotic behavior at the high energy limit (the one associated with the ideal gas of harmonic oscillators). The relevant differences appears in the low energy limit, where the non-neutral plasma approaches the Brillouin steady state, while the gravitational gas undergoes the called \emph{gravothermal collapse} and exhibits a branch with \emph{negative heat capacities}.}\label{g04_comparison.eps}
\end{figure}

The non-neutral plasma model (\ref{eq:simplified-plasma-equation}) does not exhibit microcanonical phase transitions neither negative heat capacities reported in other long-range interacting systems \cite{Gross_Book,Dauxois_Proceeding,Campa:2009PhR}. For a better understanding, let us perform a brief comparison of this non-neutral plasma model with its gravitational counterpart:
\begin{equation}\label{gravitational.model}
H(\mathbf{r},\mathbf{p})=\sum_{i}\frac{1}{2m}\mathbf{p}^{2}_{i}+\frac{1}{2}m\omega^{2}\mathbf{r}^{2}_{i}
-\sum_{i<j}\frac{G m^{2}}{\left|\mathbf{r}_{i}-\mathbf{r}_{j}\right|}.
\end{equation}
For the sake of briefly, we shall avoid to enter into mathematical details concerning the thermo-statistical analysis of this second system (the mathematical treatment is quite similar). The gravitational equivalents of Brillouin density and radius are given by:
\begin{equation}
n_{B}=\frac{3\omega^{2}}{4\pi Gm}\mbox{ and }R_{B}=\left(GM/\omega^{2}\right)^{1/3}.
\end{equation}
The gravitational radius $R_{B}$ represents the radius of the region where gravitation of this system dominates its microscopic dynamics. The gravitational radius $R_{B}$ is considered to defined the associated units for energy and temperature:
\begin{equation}
U_{B}=\frac{GM^{2}}{R_{B}}\mbox{ and }T_{B}=\frac{GMm}{kR_{B}}.
\end{equation}
We show in figure \ref{g04_comparison.eps} the temperature \emph{versus} energy dependence (the called microcanonical caloric curve) of both models using their respective characteristic units. As naturally expected, these models exhibit the same asymptotic limit for large energies (or large temperatures), where they recover the thermodynamic behavior of the ideal gas of harmonic oscillators. However, they disagree in their respective behaviors at low energies.

The repulsive character of Coulomb forces among the charge particles of the non-neutral plasma enables the existence of Brillouin steady state at zero temperature. On the contrary, the attractive character of gravitation suppress the  possibility that the model (\ref{gravitational.model}) reaches the zero temperature point. This system cannot exhibit temperatures below the critical value $T_{\beta}=0.43 T_{B}$, precisely, because of the corresponding caloric curve exhibits a minimum at the point $U_{\beta}=0.215 U_{B}$. This models also exhibit a minimum energy at the value $U_{\alpha}=0.016 U_{B}$. It is worth noticing that the heat capacity $C=dU/dT$ \emph{diverges} at the critical point $U_{\beta}$. For energies within the interval $U_{\alpha}<U<U_{\beta}$, the system temperature $T$ grows when the energy $U$ is decreased, thus evidencing the existence of a branch with \emph{negative heat capacities}. At the critical point $U_{\alpha}$, the (negative) heat capacity approaches to zero, and the system undergoes the called \emph{gravothermal collapse} \cite{Antonov1962}. According to the common understanding of this collective phenomenon (a microcanonical phase transition), the internal pressures of the system are unable to balance its own gravitational field, so that, the system undergoes a collapse that leads the formation of structures with very dense cores. The description of these post-collapse configurations, however, requires additional physical considerations beyond the classical statistical mechanics description of the model (\ref{gravitational.model}), e.g.: consideration of quantum effects, the effective linear size of constituents, etc.

Despite of the obvious differences, the non-neutral plasma (\ref{eq:simplified-plasma-equation}) and its gravitational counterpart (\ref{gravitational.model}) share many analogies. Both the Coulomb and Newtonian forces do not exhibit \emph{characteristic lengths}, so that, the stability of their thermodynamic description crucially depends on the existence of external factors, e.g.: the dynamic influence of the potential harmonic field. In particular, the thermodynamic description of both model systems depend on the frequency constant $\omega$ of the harmonic field (this constant parameter, in particular, determines the Brillouin units $R_{B}$, $n_{B}$, $U_{B}$, $T_{B}$ that enter in their low energy thermodynamic behaviors). Without the external influence of the harmonic field, the constituents will escape from both systems, thus ruling out the occurrence of a rigorous thermodynamic equilibrium. Finally, both long-range interacting systems belong to the same class of \emph{non-extensive systems}, since their observables follow in the thermodynamic limit $N\rightarrow+\infty$ the same scaling laws (\ref{next.ThL}) and  (\ref{omega.N.ThL})\footnote{Readers may object that derivation of the scaling laws (\ref{next.ThL}) and  (\ref{omega.N.ThL}) crucially depends on whether one demands or not the extensive character of the entropy. Actually, the same scaling behavior appears in the framework of the Thomas-Fermy theory for the asymptotic behaviors of atomic energy $U(Z)$ and density $\rho^{Z}(r)$ for an atom of charge $Z$ sufficiently large \cite{Thomas:1927,Scott:1952,Lieb:1977}, $U\left(Z\right) = C_{TF} Z^{7/3} + O\left(Z \right)$ and $\rho_{TF}^{Z}\left(r\right) = Z^{2}\varrho\left(Z^{1/3}r\right)$. In general, this type of scaling behavior corresponds to a three-dimensional system of non-relativistic point particles that interact among them throughout $1/r$ potential without mattering about their quantum or classical behavior.}.

\section{Results of dynamical simulations}

\subsection{Incidence of collisions}

In the framework of non-neutral plasmas, the collisions among the constituting particles is a mechanism that contributes both the chaoticity of microscopic dynamics as well as the system relaxation (at macroscopic level) towards equilibrium. If collision events constitute the main source behind the chaoticity of the present model (\ref{eq:non-neutral-hamilton}), the associated Lyapunov exponent $\lambda$ should be comparable to the frequency of collisions $\nu$ (collisions per unit of time). For the sake of convenience, let us perform a qualitative description for the present situation, which will be considered later to analyze the results obtained from the dynamical simulations. An elementary estimation is obtained in terms of the effective cross section $\sigma$ and the gas density $n$ as follows, $\nu\sim N\sigma n v_{rel}$, with $v_{rel}$ being the relative velocity among particles. The cross section $\sigma$ among charged particles can be estimated as the Rutherford scattering cross section, $\sigma\sim q^{4}/m^{2}\left\langle v^{2}\right\rangle^{2}$, with $\left\langle v^{2}\right\rangle$ being the square dispersion of velocity. Denoting the mean square velocity as $v=\sqrt{\left\langle v^{2}\right\rangle}$, the temperature dependence of both the mean square velocity $v$ the relative velocity among particles is $v_{rel}\sim \sqrt{2}v\propto\sqrt{kT/m}$.  Taking into account the asymptotic dependence of the central density $n_{0}$ for low and large temperatures ($n_{0}\sim n_{B}$ and $n_{0}\propto n_{B}\left(T_{B}/T\right)^{3/2}$, respectively), one obtains the estimation
\begin{equation}\label{rate_collisions}
\nu\propto \left\{
\begin{array}{cc}
    \omega(T_{B}/T)^{3/2}, & \mbox{ if }T/T_{B}\ll 1, \\
    \omega(T_{B}/T)^{3}, & \mbox{ if }T/T_{B}\gg 1, \\
\end{array}
\right.
\end{equation}
with $\omega$ being the frequency parameter of the external harmonic field. The previous result predict that the frequency of collisions $\nu$ decreases with the growth of temperature, but it exhibits different behaviors in the low and high temperature limits. Moreover, the characteristic timescale between collision events $\tau_{col}=1/\nu$ exhibits \emph{the same order of characteristic timescale} $\tau_{dyn}=1/\omega$ \emph{of microscopic dynamics}. Accordingly, the system as a whole will be affected by the incidence of collisions in the timescale $\tau_{c.rlx}\propto N\tau_{col}$, which represents the characteristic timescale of the \emph{collisional relaxation}.

\subsection{Initial conditions}

\begin{figure}[t!]
\centering
\includegraphics[width=4.5in]{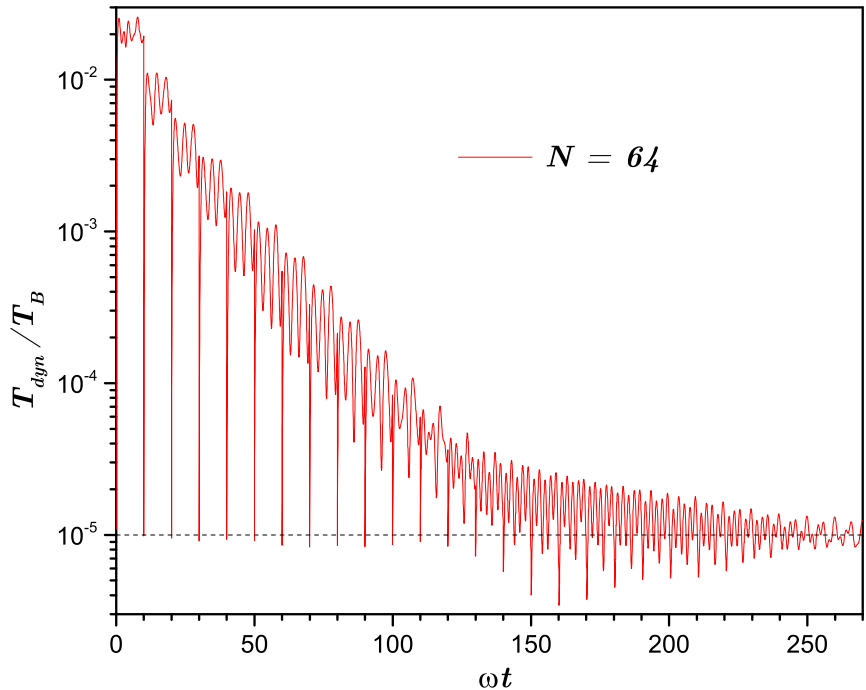}
\caption{Evolution of the dynamical temperature $T_{dyn}=2K/3N$ along the \emph{thermalization process}, which is considered here to force the equilibrium state with temperature $T=10^{-5}T_{B}$. Notice that the initial configuration significantly differs from the final equilibrium state, despite the associated temperature $T$ of the later one was indeed employed in the random generation of the initial configuration using the equilibrium distribution (\ref{MB}). Apparently, statistical fluctuations of initial positions $\left\{\mathbf{r}_{i}\right\}$ are significant to disturb the velocity distribution along the system dynamical evolution.}\label{g05_thermalization.eps}
\end{figure}

One can suppose that the model system (\ref{eq:non-neutral-hamilton}) obeys ergodicity and mixing properties that are necessary to perform a statistical description in terms of microcanonical ensemble. Besides, one can also admit that the system will arrive at an thermodynamic equilibrium by starting from an arbitrary initial condition in a finite relaxation time $\tau_{relax}<+\infty$. Since Lyapunov exponent (\ref{Lyapunov.limit}) requires infinite time limit $t\rightarrow+\infty$, this indicator will characterize the equilibrium state. In our dynamical simulations, we have started the dynamical evolution from equilibrium situation associated with Maxwell-Boltzmann distribution (\ref{MB}). Proceeding thus, we shall avoid the system initial evolution towards equilibrium and speed up all calculations.

During random generation of initial conditions using distribution (\ref{MB}), the statistical fluctuations are large for values of $N$ relatively small, overall, for low values of temperatures. In these cases, the resulting initial configuration can significantly differ from the equilibrium configuration of interest. For this reason, we have considered a \emph{thermalization} procedure to force the system to reach a given equilibrium state with temperature $T$. Along this process, the kinetic energy of each particles is periodically re-scaled in a way that the expectation value of the dynamical temperature $ T_{dyn}=2K/3N$ acquires the value $T$ of interest, $\left\langle T_{dyn}\right\rangle=T$. A particular illustration of this procedure for the system with size $N=64$ is shown in figure \ref{g05_thermalization.eps}, which was developed to force the equilibrium state with $T=10^{-5}T_{B}$.

\subsection{Verification of ergodic hypothesis}

\begin{figure}[t!]
\centering
\includegraphics[width=6.0in]{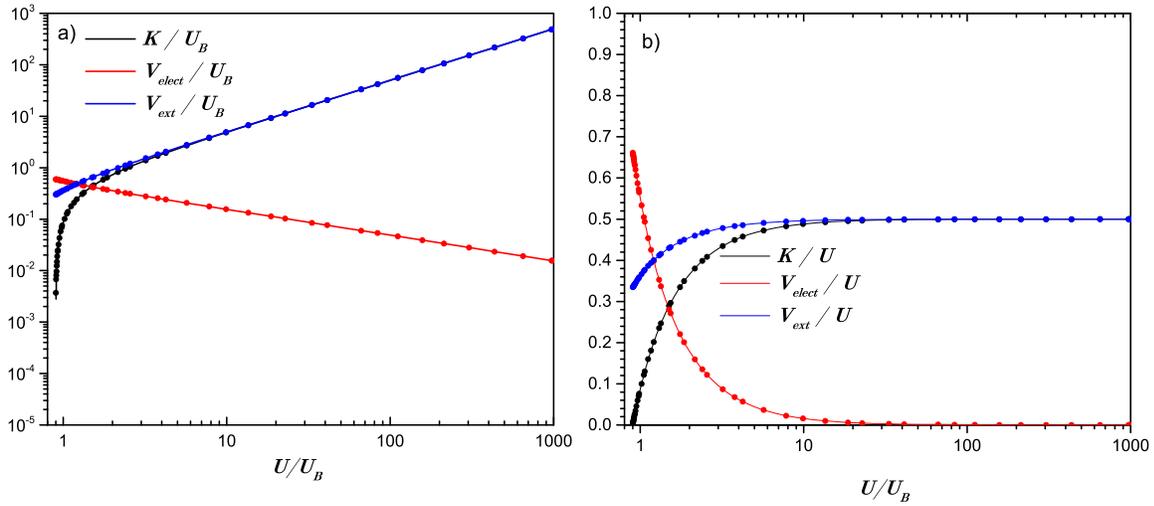}
\caption{Comparison among the microcanonical dependencies of three contributions of the total energy obtained from thermo-statistical calculations (the same results shown in Fig.\ref{g03_dyn_obs.eps}) and the temporal expectation values of these same observables obtained from dynamical simulations. The good agreement among these estimations evidences the licitness of two non-trivial considerations: i) the thermodynamic limit approximation $N\rightarrow+\infty$ and ii) the relevance of ergodic hypothesis.}\label{g04_dyn_obs.eps}
\end{figure}

Our first task is to contrast the predictions of the thermo-statistical analysis and the results obtained from the dynamical simulations. We show in figure \ref{g04_dyn_obs.eps} a direct comparison among the temporal expectation values
\begin{equation}
\left\langle O\right\rangle_{dyn}=\lim_{T\rightarrow+\infty}\frac{1}{T}\int^{T}_{0}O\left[\mathbf{q}(t),\mathbf{p}(t)\right]dt
\end{equation}
of the three contributions of the total energy $U$ (points) and their corresponding statistical expectation values (solid lines):
\begin{equation}
\left\langle O\right\rangle_{stat}=\int O\left(\mathbf{q},\mathbf{p}\right)dP\left(\mathbf{q},\mathbf{p}\right)
\end{equation}
already shown in figure \ref{g03_dyn_obs.eps}. The agreement of these predictions is excellent taking into consideration that our dynamical simulations were restricted to a number of particles $N=2048$, while the statistical estimations were performed by invoking thermodynamic limit $N\rightarrow+\infty$. The present results evidence (1) the applicability of the thermodynamic limit approximation $N\rightarrow+\infty$ to obtain statistical expectation values for finite systems with moderate number of particles $N\sim 10^{3}$ and (2) the \emph{ergodic character} of the non-neutral plasma model (\ref{eq:simplified-plasma-equation}) considered in the present study, namely, the equalization between the statistical and dynamical expectation values of a given observable $O$:
\begin{equation}\label{ergodic}
\left\langle O\right\rangle_{dyn}=\left\langle O\right\rangle_{stat}.
\end{equation}

\begin{figure}[t!]
\centering
\includegraphics[width=4.5in]{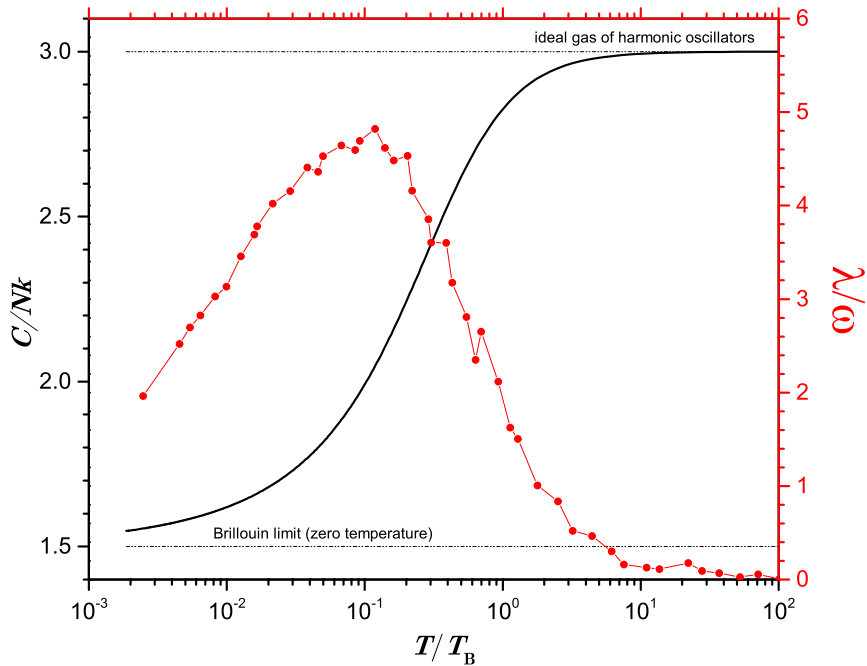}
\caption{Comparison between the heat capacity per particles $C/Nk$ and the Lyapunov exponent $\lambda/\omega$ for $N=2048$ (in units of the typical time $\tau_{c}=1/\omega$) \emph{versus} the temperature $T$ (in units of Brillouin temperature $T_{B}$). Accordingly, the system chaoticity reaches a maximum along the transition between the two asymptotic limits of the non-neutral plasma model (\ref{eq:simplified-plasma-equation}). The observed temperature dependence of the Lyapunov exponent cannot be explained by the rate of particles collisions (\ref{rate_collisions}). In fact, such a chaoticity is associated with the non-linearity of microscopic dynamics due to the presence of Coulombian inter-particles forces, which must reach its maximum along the transit between the two asymptotic limits of this model.}\label{g05_liapunov.eps}
\end{figure}

\begin{figure}[t!]
\centering
\includegraphics[width=4.5in]{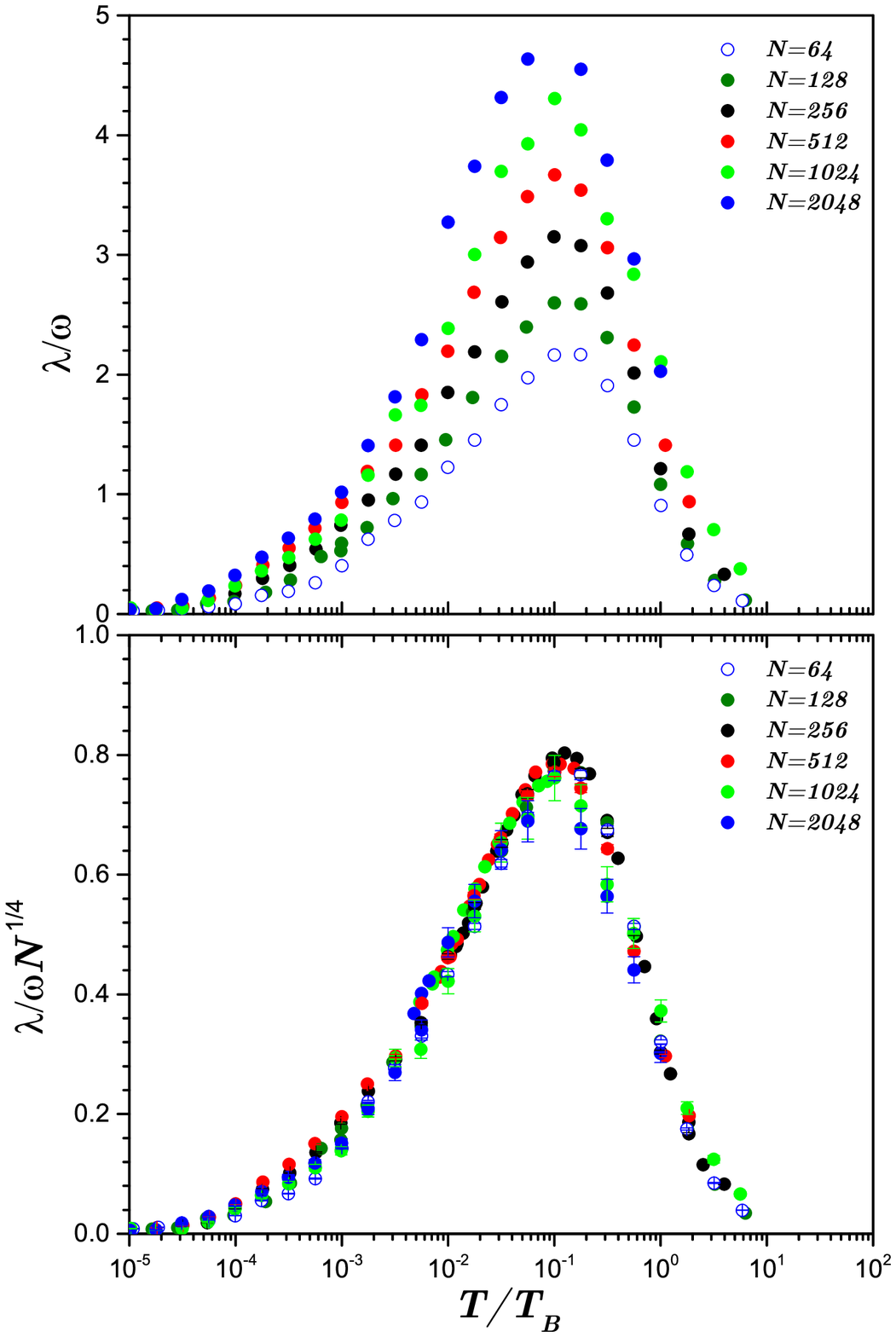}
\caption{Top panel: Size effects in the dependence of Lyapunov exponent \emph{versus} temperature obtained from extensive simulations in the range $N=64-2048$. Bottom panel: The same dependencies re-scaled by the factor $N^{1/4}$, where one observes a good confidence in the existence of a \emph{data-collapse} under the statistical uncertainty of these calculations (error bars are also added).}\label{g06_chaos_size.eps}
\end{figure}

\subsection{Chaoticity of the microscopic dynamics}

Let us now analyse the relationship between the chaoticity of the microscopic dynamics and its thermodynamic behavior. Specifically, we have performed extensive simulations to obtain the dependence of Lyapunov exponent $\lambda$ on the system temperature $T$. Results for a number of particles $N=2048$ are shown in figure \ref{g05_liapunov.eps}. In our simulations, the time variable $t$ was referred to into units of $\tau_{c}=1/\omega$, and hence, the Lyapunov exponent $\lambda$ (the maximum one) is expressed to in units of constant frequency $\omega$ of the external harmonic field. For comparative purposes, we have also included in this figure the dependence of heat capacity $C$ \emph{versus} temperature $T$ already shown in figure \ref{g01_limits.eps}. According to these results, the Lyapunov exponent decreases (and probably drops to zero) when the system approaches its two asymptotic limits of low and high temperatures, while it exhibits a local maximum during the transition around the temperature value $T\simeq 0.1 T_{B}$. The observed temperature dependence of the Lyapunov exponent differs from the one considered by the rate of collisions (\ref{rate_collisions}). Of course, one cannot expect a direct identification between these quantities, but the growth of Lyapunov exponent in the low temperature limit cannot be explained in terms of the particles collisions. The mechanism of collisions turns more effective for low temperatures, which is in contradiction with the reduction of the system chaoticity observed when the temperature decreases.

For a better understanding of the system chaoticity, we have studied the dependence of the Lyapunov exponent $\lambda$ on the system size $N$. Results of extensive simulations considering the ranges $N=64-2048$ and $T/T_{B}=\left[10^{-5}-10\right]$ are shown in figure \ref{g06_chaos_size.eps}. According to these results, the Lyapunov exponent exhibits the same qualitative dependence on the temperature $T$, but the overall values of this quantity grow with the system size $N$. A simple analysis evidences that the Lyapunov exponent $\lambda$ (in units of $\omega$) grows with $N$ following the power-law $N^{1/4}$. This scaling law is evidenced by the data-collapse after re-scaling $\lambda/\omega$ by the factor $N^{1/4}$ (the bottom panel of figure \ref{g06_chaos_size.eps}).

The present results evidence that the characteristic chaotization timescale of this non-neutral plasma model is $\tau_{ch}=1/\lambda\propto 1/\omega N^{1/4}=\tau_{dyn}/N^{1/4}$, which is \emph{considerably smaller than the characteristic timescale of its microscopic dynamics} $\tau_{dyn}$, $\tau_{ch}\ll \tau_{dyn}$. The observed chaoticity cannot be explained in terms of particles collisions because of their characteristic timescales considerably differ between them. The size dependence of the Lyapunov exponent should be explained by some type of collective influence of the system as a whole. Taking into consideration precedent studies in the context of astrophysical model \cite{Cerruti-Sola1995a,Cipriani2003}, the observed chaoticity should be explained in terms of the phenomenon of \emph{parametric resonance}. The verification of this hypothesis requires the application of Riemannian approach of Hamiltonian chaos \cite{Pettini1993}, which is beyond the scope of the present study.

The chaoticity of a Hamiltonian system is understood as a consequence of non-linearity of its microscopic dynamics. Apparently, the incidence of non-linear effects in this concrete situation reaches its maximum during the transition between the two asymptotic limits of the system thermodynamic behavior. According to the heat capacity $C$ \emph{versus} temperature dependence, during the transit between Brillouin limit towards the ideal gas of harmonic oscillators limit, it takes place the \emph{unfreezing} of the oscillatory degrees of freedom. Therefore, the non-linear effects of microscopic dynamics that explain the chaoticity shown in figures \ref{g05_liapunov.eps} and \ref{g06_chaos_size.eps} must be in someway associated to this process of unfreezing of the oscillatory degrees of freedom.

\section{Final remarks and open questions}

We have studied in this work the thermodynamics and the dynamics of a simple model of a pure non-neutral plasma confined under an external harmonic field,  Eq.(\ref{eq:simplified-plasma-equation}). Despite its simplicity, this model preserves essential features of more realistic models of non-neutral plasmas confined in magnetic traps, like the Brillouin steady state, as well as the non-extensive character due to the long-range character of Coulombic forces among charged particles. According to results obtained during dynamical simulations, the observed chaoticity of the present model is very strong, since it take places at a rate faster than the characteristic timescale $\tau_{dyn}$ of the microscopic dynamics. According to our preliminary qualitative analysis, such a strong chaoticity cannot be explained in terms of collision events because of their respective characteristic timescales are significantly different. In fact, such a strong chaoticity is the result of some type of collective phenomena that attains the system as a whole, presumably the called resonance parametric mechanism proposed by Pettini and co-workers as the \emph{main source} of chaoticity in the context of nonlinear Hamiltonian systems with bound motions in the configuration space \cite{Pettini1993}.

In accordance with the chaotic hypothesis \cite{GallavottiCohem:1995}, the strong chaoticity observed in this model suggests the relevance of statistical properties like ergodicity and mixing for pure non-neutral plasmas. This idea is in someway corroborated in our numerical simulations shown in figure \ref{g04_dyn_obs.eps}, which evidenced the good agreement of thermo-statistical calculations and its associated temporal expectation values. Certainly, the present situation is not subjected to evaporation events as the case of the experiment of Huang and Driscoll in the past \cite{Huang1994}. Nevertheless, the strong chaoticity in non-neutral plasmas should not significantly depend on the particles evaporation, but on the long-range character of Coulombian forces. Results obtained in this work reinforces the licitness of \emph{effective quasi-ergodicity} invoked in the precedent study developed by Ordenes-Huanca and Velazquez \cite{Ordenes-Huanca2016}, and why their theoretical development provides a good characterization of the experimental profiles reported by Huang and Driscoll (where it was a negligible incidence of collision events).

Let us finally refer to the open problems of this work. Firstly, the present model should be considered within the Riemannian approach of Hamiltonian chaos \cite{Pettini1993} in order to check if the parametric resonance is origin of the observed strong chaoticity. Secondly, a way to check the connection between the Lyapunov exponent and the effective unfreezing of the oscillatory degrees of freedom is to attempt the analytical computation of Lyapunov exponent using the ideas of Casetti and Pettini \cite{Casetti1993}. Thirdly, it is necessary to check the incidence of strong chaoticity of this model on its relaxation time towards thermodynamic equilibrium. The possible relation between these two timescales is someway suggested by the known relation of Sinai–Komolgorov entropy $h_{KS}$ (a measure of the entropy production) and the sum of all positive Lyapunov exponents \cite{Pesin1977}:
\begin{equation}\label{KS}
h_{KS}\sim\frac{dS_{KS}}{dt}\equiv\sum_{i}\lambda_{i}^{+}\Rightarrow \tau_{rel}\propto \tau_{ch}N,
\end{equation}
where $N$ is the size of the system, while $\tau_{ch}$ and $\tau_{rel}$ are the chaotization and relaxation timescales, respectively. If this heuristic relationship between these two timescales is correct, its existence would explain the relevance of chaoticity on the ergodicity of microscopic dynamics (the effective filling of the energy surface in the phase space).

\section*{Acknowledgments}

Authors thank partial financial support of this research from FONDECYT \textbf{1170834} and CONICYT-PAI \textbf{79170075} (Chilean agency).

\newpage
\section*{References}
\bibliographystyle{iopart-num} 
\bibliography{plasma}

\end{document}